\documentclass[doublespacing]{elsart}

\usepackage{graphicx}
\usepackage{amssymb}

\newcommand{\CDMS}{{\small CDMS}}

\newcommand{\CDMSII}{{\small CDMS\,II}}
\newcommand{\WIMP}{{\small WIMP}}
\newcommand{\WIMPs}{{\small WIMP}s}

\begin{document}

\begin{frontmatter}

\title{A Neutron Multiplicity Meter for Deep Underground Muon-Induced 
High Energy Neutron Measurements}

\author{R. Hennings-Yeomans\corauthref{cor1}} and
\corauth[cor1]{Corresponding author: R Hennings-Yeomans}
\ead{raul@casino.phys.cwru.edu}
\author{D.S. Akerib}
\address{Case Western Reserve University, Department of Physics, 10900 Euclid Ave, Cleveland OH 44106-7079}

\begin{abstract}
We present, for the first time, the design of an instrument capable of measuring 
the high energy ($>$60\,MeV) muon-induced neutron flux deep 
underground. The instrument is based on applying the Gd-loaded 
liquid-scintillator technique to measure the rate of multiple low energy neutron events 
produced in a Pb target and from this measurement to infer the rate of high energy neutron 
events. This unique signature allows both for efficient tagging of neutron muliplicity 
events as well as rejection of random gamma backgrounds so effectively that 
typical low-background techniques are not required. We present design 
studies based on Monte Carlo simulations that show that an apparatus 
consisting of a Pb target of 200\,cm by 200\,cm area by 60\,cm thickness covered by a 
60\,cm thick Gd-loaded liquid scintillator (0.5$\%$ Gd content) detector 
could measure, at a depth of 2000\,meters of water equivalent, a rate of $70\pm8$ (stat)\,events/year 
with a background of less than 10\,events/year. We discuss the relevance of this 
technique to measuring the muon-induced neutron background in searches for dark 
matter in the form of  Weakly Interacting Massive Particles. Based on these studies, 
we also discuss the benefits of using a neutron multiplicity meter as a  component of 
active shielding in such experiments.

\end{abstract}

\end{frontmatter}
\newpage
\section{Introduction}
\label{motivation}
The nature of dark matter is one of the most important outstanding
issues in particle physics, cosmology and astrophysics. A leading
hypothesis is that Weakly Interacting Massive Particles, or \WIMPs,
were produced in the early universe and make up the dark matter.  So
far this matter has only been observed through its gravitational
effects. \WIMPs\ cannot be Standard Model particles and so their
discovery would hail a new form of matter. A detection would also help
solve a long-standing riddle in cosmology that even questions our
understanding of gravity.  Dark matter is concentrated in the halos of
galaxies, including the Milky Way. If \WIMPs\ make up these halos they
can be detected via scattering from atomic nuclei in a terrestrial
detector.  Experiments that search for \WIMPs\ are one of the critical
science drivers for a Deep Underground Science and Engineering
Laboratory in the United States. 

\WIMP\ searches must be performed underground to shield from 
cosmic rays, which produce secondary particles that could fake a \WIMP\ signal.
Nuclear recoils from fast neutrons in underground laboratories are one
of the most challenging backgrounds to \WIMP\ detection. Experiments 
that search for \WIMP\ dark matter rely on passive and active shielding to 
reduce gamma and neutron backgrounds. To reduce
the neutron background, passive hydrogen-rich shielding and active
charged-particle detectors are commonly used to moderate neutrons and
veto muon-induced events, respectively.  To reduce the gamma
background, high-$Z$ materials such as lead are used to attenuate
gammas from ambient radioactive sources. While the high-$Z$ shielding
is effective against gammas, the shield itself becomes a source of
increased neutron background due to secondary particles produced by
unvetoed muon-induced neutrons that have energy above about 60\,MeV.
These neutrons have sufficient energy and low enough cross section on
hydrogen that they penetrate the moderator and reach the gamma
shield. They tend to interact there and cause spallation reactions,
which produce multiple secondary neutrons with energy below
10\,MeV. At these lower energies, the neutrons can reach the inner
detector volume and cause \WIMP-like nuclear recoils.

The high-energy neutrons and their parent reactions that originate
with cosmic-ray muons are thus correlated with the unvetoed neutron
events that mimic the \WIMP\ signal. Neutron production by muons
underground have been measured at a span of depths and muon energies,
from about 20 meters of water equivalent (\,m.w.e.) depth and 10\,GeV 
energy~\cite{25d1, 25d2, Bezrukov} to 5200\,m.w.e. and 400\,GeV~\cite{Mont1}. 
An estimate of the neutron production as a function of muon energy for muons 
interacting in liquid scintillator has been obtained by Wang and
co-workers~\cite{Wang1} based on Monte Carlo simulations made with 
{\small FLUKA}~\cite{FLUKA0} that is about a factor of two 
within the available data at that time for depths with a mean muon energy 
above 100\,GeV. Since we are interested in neutrons with energy above $\sim$60\,MeV, 
we note this work is primarily sensitive to neutrons below this energy range 
as illustrated by Figure~5 in their paper~\cite{Wang1}. Galbiati and 
Beacom~\cite{GalbiatiAndBeacom} have calculated, using {\small FLUKA}~\cite{FLUKAGB}, 
the production rates for $^{12}$B in muon--induced showers and have probed 
the neutron production in the energy range of $\sim$10-100\,MeV through 
the $^{12}$C(n,~p)$^{12}$B reaction and the calculation agrees well with 
measurements of $^{12}$B at 2700\,m.w.e. made by KamLAND~\cite{kamland}. 

At higher energies ($>$100\,MeV) the shape of the neutron 
spectrum was compared to {\small FLUKA}~\cite{FLUKA0} by Wang~\cite{Wang1} 
and there is about a factor of two discrepancy with data taken with the 
liquid-scintillator {\small LVD} detector~\cite{lvd} at a depth of 3650\,m.w.e 
and a mean muon energy of 270\,GeV. Mei and Hime~\cite{MeiHime} claim 
that after making corrections for proton recoil quenching effects, the corrected 
{\small LVD} data agrees well with the shape of the spectrum predicted with {\small
FLUKA} simulations. However, the individual who performed the analysis of the data, 
V.A. Kudryavtsev has pointed out that this correction is inappropriate~\cite{Kpersonal}. 
At present, there is no other data at this energy to inform the production of high energy 
neutrons, leaving the discrepancy unresolved. The LVD collaboration recently presented 
results on measuring neutron production above 20\,MeV by muons in liquid 
scintillator~\cite{ICRC2005}, and a Monte Carlo simulation is under development by 
the collaboration to convert this measurement to an absolute flux.

Neutron production by 200\,GeV muons occurs through hadronic
showers generated by the muons interacting in the rock, and to some
extent by direct muon spallation~\cite{Wang1}. The {\small CERN} {\small NA55}
experiment measured neutron production via direct muon spallation by
looking at the production of fast neutrons ($>$10\,MeV) by 190\,GeV
muons on graphite, copper and lead~\cite{Chazal} at three different
angles from the muon beam.  Araujo and co-workers~\cite{Araujo} show
that this experimental data lies above the Monte Carlo simulations
from between a factor 3 to 10 depending on the measured angle. These 
measurements could overestimate the rate because of  contamination by
neutrons produced by secondaries of the muon-nucleus interaction. 
The possible systematic uncertainties leave the
matter inconclusive, informing neither muon spallation production nor
the total fast neutron yield above $>$10\,MeV.

The measurements to date of neutrons at large depths 
involve either primary muon interactions in hydrocarbon
liquid scintillator followed by cascade processes within 
the detector~\cite{lvd, kalchukov}, or muon
interactions in higher-$Z$ material such as Pb and Cu~\cite{gorshkov}
in which neutron production is dominated by relatively low-energy
electromagnetic properties. Of particular interest for dark
matter experiments, as noted above, is when high-energy neutrons produced in the
rock through muon interactions and hadronic cascades, followed by
spallation in high-$Z$ shielding, lead to a flux of neutrons of mostly 10\,MeV
and below. In the work described here our simulations and calculations
indicate that a modest size detector, by exploiting the multiplicity
distribution of the spallation events, can provide a normalization of
the neutron flux to a precision of about 12\%. By measuring
the high energy neutron flux at 2000\,m.w.e. we will benchmark the
neutron production by muon induced hadronic showers and provide a
normalization of the unvetoed neutron background. We have chosen this 
depth because the muon-induced neutron production is dominated by hadronic processes 
according to Wang and co-workers~\cite{Wang1} based on Monte Carlo simulations made with 
the particle production and transport code {\small FLUKA}~\cite{FLUKA0} 
and because the rate is good enough for a modest detector size (see section~\ref{Principle}) 
to be able to measure a rate of $70\pm8$\,events/year for neutrons $>$60\,MeV.

In addition to the interest for the shielding configurations for many
dark matter experiments, improved knowledge and predictibility of the muon-induced 
high energy neutron flux ($>$60\,MeV) at depth will aid in the understanding of neutron induced
backgrounds in double beta decay experiments. For example, as noted by
Mei and Hime~\cite{MeiHime}, knowledge of the neutron background is
needed to estimate the background due to elastic and inelastic events
that generate gamma rays near the 2\,MeV endpoint, and to optimize
shielding configurations that also typically involve massive lead and
polyethylene shields to attenuate gammas and moderate neutrons. Thus
for two major classes of low-background underground experiments, dark matter and double
beta decay, a more precise measurement of the neutron background
produced in the appropriate shield components will be of great utility, 
from the experiment planning stage through to data analysis.

\section{Principle of the Instrument}
\label{Principle}
The instrument we have designed is based on applying the Gd-loaded
liquid-scintillator technique to measure the rate of events with multiple low 
energy neutrons produced in a Pb target.  Our studies, which are presented in 
Section~\ref{design}, indicate that at a depth of 2000\,m.w.e., the dominant 
source of these events is due to muon--induced high energy neutrons interacting 
in the Pb. Gadolinium has a high thermal-neutron capture cross section, and 
emits 8\,MeV in gamma rays after the capture. Since neutrons thermalize and capture 
with a mean of about 10\,$\mu$s, measurements of the distinct capture times is a straightforward 
way to determine neutron multiplicity, and to tag and measure the underlying process of
the fast-neutron production.  This method, known as a Neutron Multiplicity Meter, has a 
long history of use, dating to searches for superheavy elements expected to decay to 
high-neutron-muliplicity final states~\cite{SSEN}, and more recently in accelerator-based
applications~\cite{BBN}.

\begin{figure}[htbp]
\begin{center}
\includegraphics[width=6in]{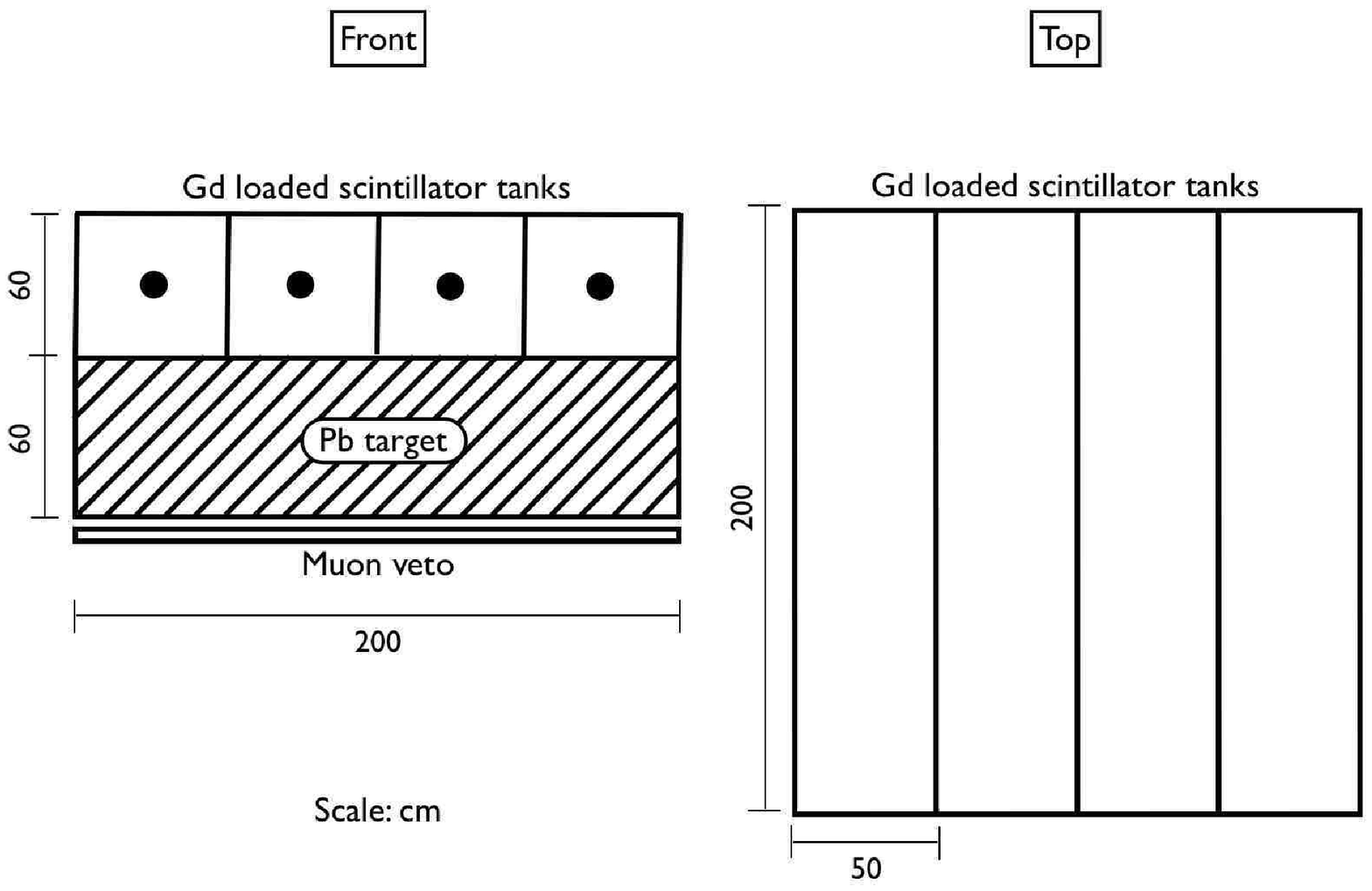}
\caption{\small Conceptual drawing of the Neutron Multiplicity Meter for Deep 
Underground Muon-Induced High Energy Neutron Measurements. 
}
\label{fig:fig_d3}
\end{center}
\end{figure}

The basic design of the Neutron Multiplicity Meter 
applied to measure high energy neutrons ($>$60\,MeV) underground
employs the Gd-loaded liquid-scintillator detector ($\sim$0.5$\%$ Gd content) atop a 
200-cm-square by 60-cm-thick Pb target in which
high energy neutrons produced by muon interactions in the rock
walls of the cavern will mainly enter from above, penetrate the
scintillator, and cause neutron spallation in the Pb, as illustrated in Fig.~\ref{fig:fig_d3}. 
The secondary low energy neutrons produced by the primary high energy neutron 
leave the Pb target and enter the Gd-loaded scintillator, where they are moderated and 
thermalized by the protons in the hydrocarbon which comprises the bulk of the
scintillator. Within about 40\,$\mu$s, most will have captured on
the gadolinium, and thus the essential problem of detecting neutral
particles with high efficiency has been turned to an advantage: the
neutrons which are released simultaneously 
are dispersed in time, and individually captured and counted.  As the
simulations below illustrate, this unique signature allows both for
efficient tagging of neutron muliplicity events as well as rejection
of random gamma backgrounds so effectively that typical low-background
techniques are not required. 

\section{Instrument Design Studies}
\label{design}
\begin{figure}[htbp]
\begin{center}
\includegraphics[width=4in]{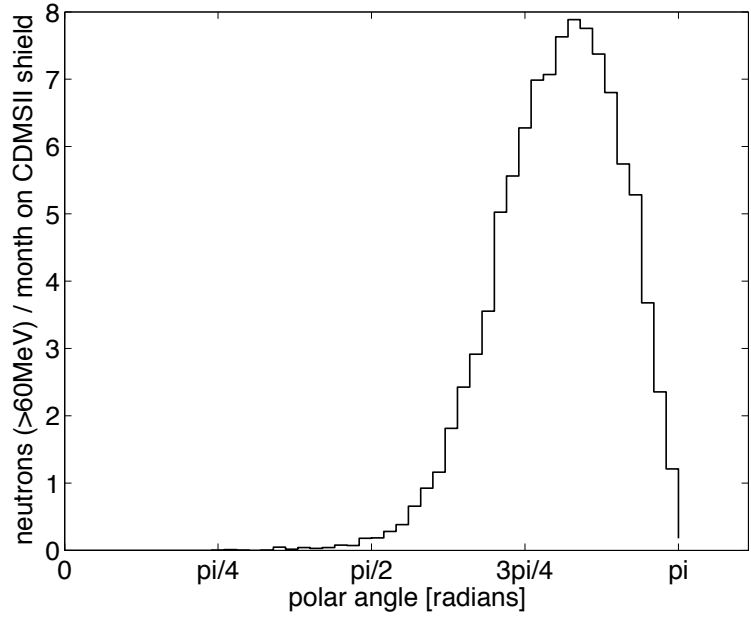}
\caption{\small Polar angular distribution of the neutrons with energy
greater than 60 MeV incident on the CDMSII shield. Neutrons tend to
have downward direction at an angle of about 0.88$\pi$ radians with respect
to the normal vector from the floor.  Therefore the area of the target
is proportional to the rate of incident high energy
neutrons.  }
\label{fig:hencosz}
\end{center}
\end{figure}

In this section the design characteristics of the Neutron Multiplicity
Meter adapted to measure high-energy neutron flux underground are
developed. Extensive simulation studies of the muon-induced neutron
background in the Soudan Mine at a depth of 2000\,m.w.e. corresponding
to 14 years of exposure have been performed using 
{\small FLUKA} simulation package~\cite{FLUKA1,FLUKA2}. These studies, 
carried out for background estimates in the \CDMSII\ experiment, are based 
on an angular distribution of muons matched to this depth, and 
normalized to the measured flux in the \CDMSII\ plastic-scintillator veto 
system~\cite{r118PRD}. In the study, the muons are propagated into a rock-wall cavern modeled as a
6-sided 10-m-thick rock shell surrounding a 4\,m by 8\,m by 4\,m
cavity.  The \CDMSII\ experimental setup is inside the cavity and near
one of the walls. High energy neutron production due to muons occurs
through direct muon spallation and subsequent hadronic showers that
develop in the rock.  The angular distribution of neutrons above 60
MeV, as depicted by the distribution in Figure~\ref{fig:hencosz}, shows
that the neutrons are mostly going downward at angles greater than
about 0.88$\pi$\,radians, where $\pi$ radians corresponds to the direction 
vertically downward. Given the predominantly downward direction, the rate of
incident high-energy neutrons is proportional to the area of the Pb
target, which defines the first criterion for the setup.

\begin{figure}[htbp]
\begin{center}
\includegraphics[width=4in]{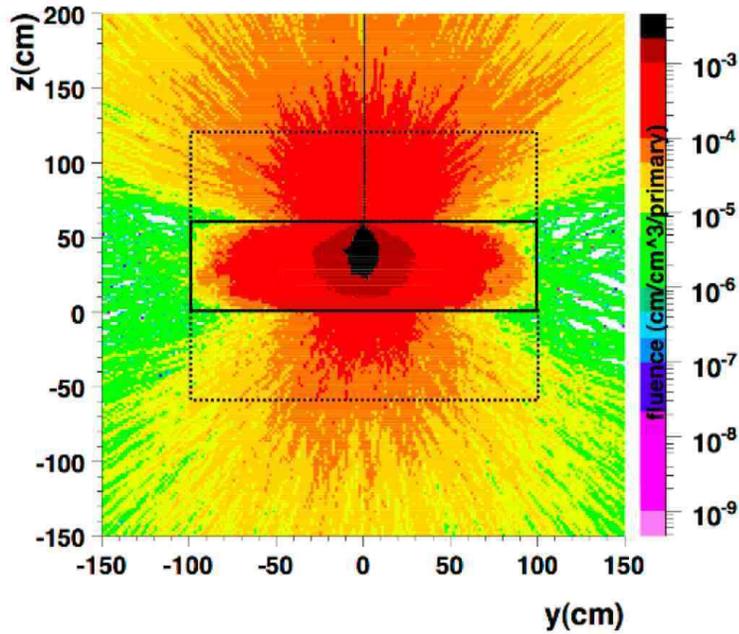}
\caption{\small Neutron fluence plot of a FLUKA simulation that propagates
a beam of 100\,MeV neutrons on a 60-cm-thick Pb target. The upper and
lower rectangles are reference surfaces delimiting the counting
boundary for the upper and lower neutrons, respectively. The central
rectangle from $z$=0 to 60 is the Pb target.  The plot shows more
evaporated neutrons going upwards than downwards (the forward
direction relative to the beam) due to backscattering. This effect 
also causes very few neutrons to go forward as the thickness increases
above about 20 cm.  }
\label{fig:fluence}
\end{center}
\end{figure}

The next criteria we consider for the Pb target are the optimal thickness
and whether it is best placed above or below the scintillator tank. A
simulation with {\small FLUKA} was performed by propagating a beam of
100-MeV neutrons at a 200\,cm by 200\,cm Pb target with thickness varying
from 1 to 100\,cm. We gauge the detectability of a subsequent
multiplicity event by counting the number of secondary neutrons that
emerge from the Pb with less than 10\,MeV and are thus readily
moderated and captured. We define the parameter $P$ for both the top
and bottom surfaces as the fraction of events for which a
downward-direction 100-MeV neutron results in at least 3 low-energy
neutrons exiting either the side from which the neutron beam was incident 
(top) or the opposite side (bottom). The overall production point and
neutron travel direction is illustrated in Figure~\ref{fig:fluence},
which shows the neutron fluence (neutron track length per unit volume)
in units of cm per cm$^3$ per primary neutron, based on a
{\small FLUKA} simulation for a 60-cm-thick target. Quantitative results 
for $P$ are shown in Figure~\ref{fig:flipped_or_NOT}, where the ``Pb target 
on bottom" means $P$ is calculated for downward incident neutrons with
updward-going secondaries to be detected in a top-side scintillator
detector, and ``Pb target on top" means $P$ is calculated for downward
secondaries to be detected bottom-side. 

We observe that the emission of neutrons is roughly isotropic as expected, 
and that the spallation reaction occurs within the first 15\,cm of
Pb. Furthermore, as the thickness of the target increases beyond
20\,cm, more of the secondaries are going upwards than downwards. 
This effect is due to backscattering from the Pb, which acts roughly like a 
``neutron mirror" for low energy neutrons, since the elastic collisions off the 
Pb nuclei do little to reduce the energy of the comparatively light neutrons. 
Most important for the overall configuration, we see that since the primary 
interaction rate is still increasing with thickness, the backscatter effect 
indicates that the multiplicity rate is higher on the top side, and higher 
for increasing thickness. To maximize the detected multiplicity rate, it is 
better to place the scintillator atop the Pb, which also has the advantage of 
tagging muons that strike the Pb directly.

So far, the detector configuration is to have the Gd-loaded scintillator on 
top of the Pb target. Since neutrons with an energy less than about 60\,MeV 
will scatter off the protons in the scintillator, they will tend to either fail to reach 
the Pb or reach it with insufficient energy to produce a multiplicity of 3 or more.
In other words, the scintillator will filter low energy neutrons and together with the 
requirement that the event has a multiplicity threshold of 3 or more secondary neutrons 
this will select only those primary neutrons with an incident energy of 60\,MeV or more. 
To illustrate that high energy neutrons ($>$60\,MeV) induce a multiplicity of 3 or more 
low energy neutrons on a Pb target, Figure~\ref{fig:mult_dist} shows the induced 
detectable multiplicity, for a geometry of the Pb to have an area of 200\,cm by 200\,cm 
normal to the vertical, a thickness of 100\,cm, and an incident downward-going neutron 
beam in {\small FLUKA}  at energies of 60, 100, and 200\,MeV. The detectable multiplicity 
was estimated by counting the number of neutrons below 10\,MeV that enter a top-side 
detector with the same footprint as the Pb. The resulting multiplicity distributions for the
three energies are shown in Figure~\ref{fig:mult_dist}, where ``Event
Fraction" corresponds to the fraction of events with respect to the total number 
of incident neutrons.  The plot shows that the majority of the events have a detectable 
multiplicity of 3 or more, and that there is an increase in multiplicity with primary neutron energy 
and although some information on the primary neutron energy is potentially available 
from the multiplicity distributions; at least an energy threshold on the primary neutron energy 
can be established using multiplicity, which has a fairly sharp turn on at 60\,MeV for a 
multiplicity threshold of 3.
 
It is important to estimate the efficiency of the selection criteria 
for tagging high-energy neutron events as a function of multiplicity 
so that an optimization can be made to reject random coincidences and 
still achieve good efficiency for neutron-induced events. We identify a 
class of ``clean'' multiplicity events, that is, those that are clearly produced 
by high energy neutrons interacting in the target as opposed to other 
charged particles or gamma rays that may also have been produced by 
the parent muon. To estimate the rate of these events as a function of multiplicity we  
use the events with neutron energy above 60\,MeV from our 14--year Soudan  
simulation in which associated gamma rays,  
muons, or hadrons deposit less than 2\,MeV in the scintillator. The  
multiplicity is counted by considering only those secondaries with  
energy less than 10\,MeV entering a top-side detector, and is plotted  
in Figure~\ref{fig:anal_multiplicity_60cm}. To see the effect of 
tightening the multiplicity cut to reduce the probability of random 
coincidences, the integral number of multiplicity-tagged events per 
year is plotted versus the minimum required multiplicity, and is 
displayed in Fig.~\ref{fig:anal_effmult}. The total number of events changes only 
by about 10\% between a minimum multiplicity of 3 and 10.  

In determining the optimal thickness of the scintillator modules, we
consider two requirements: the moderation of the secondary neutrons,
and the absorbtion of the Gd capture gammas. The {\small FLUKA}
simulation predicts that the spectrum of neutrons emerging from the Pb
falls off almost completely by 5\,MeV, as shown in
Figure~\ref{fig:anal_fluxspec}. A scintillator region of 10\,cm
thickness would be sufficient to moderate them. However, we find that
containing the capture gammas requires a thicker detector.  In order
to find the optimal thickness, we used the low-energy simulation code,
{\small MCNP}-PoliMi~\cite{mcnp-polimi}, which includes the neutron-capture
process. A beam of 0.5-MeV neutrons was propagated from the Pb up to a
top-side scintillator tank, and the thickness of the tank was
varied. In Fig.~\ref{fig:scint_thickness} the efficiency to detect the gamma cascade 
with a 3-MeV threshold is shown as a function of scintillator
thickness. To allow for resolution effects, we choose 3\,MeV as the
nominal lower analysis threshold to gain immunity from gammas from
natural radioactivity, the highest of which comes from $^{208}$Tl with
an energy of 2.6\,MeV. We find that the detection efficiency increases 
with thickness because of improved containment of the gamma cascade.
The efficiency to detect 3\,MeV energy depositions from gamma-rays in 
the Gd-loaded scintillator tanks is considered to be 100\%, as this can be 
easily achieved with a 5" PMT for the configuration shown in Fig.~\ref{fig:fig_d3}.

To assess the rate of background coincidences that can mimic the
signal, we consider not just the energy criteria of nominally
3--8\,MeV for individual captures, but also the time distribution of
the captures.  The time profile for the moderation, thermalization, diffusion and
capture of multiple neutrons released simultaneously into the
scintillator is broad, with a peak at about 10\,$\mu$s after
emission and about 90\% of captures ocurring within the first
30\,$\mu$s.  It is the diffusion of the neutron what dominates the 
time between moderation and capture. A neutron burst results in a 
cleanly-separated readily-counted pulse train since the pulse widths 
of about 10\,ns are narrow compared to the typical time between captures of order
1\,$\mu$s.

Ambient gamma rays, which dominate the rate of random events in the
detector, can mimic a high energy neutron event due to accidental
coincidences within the time and energy window defined for
multiplicity events. The rate of gamma-induced background as a function of
the multiplicity criterion is shown in Fig.~\ref{fig:gamma_bkg}
for a time window of 40\,$\mu$s and three different gamma rates.
The gamma rate at Soudan expected in the Gd liquid
scintillator volume is about 600 Hz, based on gamma rates measured
with the \CDMSII\ plastic scintillator panels for a 1\,MeV
threshold~\cite{Joel}.  A reduction of an order of magnitude in rate
can be achieved with a threshold of 3\,MeV, which will render the rate
of accidental 3-fold multiplicity events to $10^{-2}$ per day, or
about one order of magnitude below the multiplicity rate predicted
from high-energy neutrons interacting in the Pb. Further reduction of 
the gamma ray rate can be achieved, if necessary, with a thin layer of 
Pb surrounding the scintillator. Alternatively, immunity from random 
coincidences can be gained by increasing the multiplicity criterion. 

We also consider the background due to neutrons from radioactivity,
which are dominated by alpha-n reactions in the rock originating from
alpha decays in the uranium and thorium decay chains.  The ambient rate of
neutrons from radioactivity at Soudan is estimated from the
measurements of the U/Th contamination in the Soudan
rock~\cite{Ruddick} and cross referenced with measurements of
both the U/Th level and neutron flux at the Kamioka
mine~\cite{Kamioka}. The resulting flux estimate of about $2 \times
10^{-5}$ neutrons/cm$^{2}$/sec produces a rate of about 3 neutrons/sec
in a detector with a scintillator volume of 200$\times$200$\times$60\,cm$^{3}$, 
and is a negligible source of multiplicity events.

Spontaneous fission from the $^{238}$U in the rock could in principle
produce events with multiplicity of 3 or more, although the most
frequent multiplicity is 2. However, the relative rate of fissions to
gammas from $^{238}$U in secular equilibrium is down by 6 orders of
magnitude. If the entire rate of ambient gammas is attributed to
$^{238}$U, the expected rate of multiplicity events from
fission would still be negligible. However, if needed, a layer of
10--20\,cm of polyethylene can easily shield them. 

Events in which the muon itself passes through the scintillator are 
also considered. Most minimum ionizing muons will have sufficient 
pathlength of about 5\,cm in the scintillator to be readily distinguished 
from Gd capture, allowing us to study muon-tagged events. For example, 
some of these muons will interact directly in the Pb, and produce a 
detectable population of neutron multiplicity events. While these events 
are of interest, they are dominated by low energy electromagnetic 
processes~\cite{kalchukov} and so are not as useful a cross check on 
the unvetoed population, which is dominated by higher-energy hadronic 
processes.

These tagged muon events will be identified by requiring more than 9\,MeV 
in the scintillator, that is, above the maximum that can be 
caused by a neutron capture. However, this criterion will also 
include some events with no muon in the scintillator but which have instead 
a high-energy neutron that deposits more than 9\,MeV by scattering in 
the scintillator. Based on a FLUKA simulation, the fraction of high-energy 
neutrons impinging on the apparatus that are in this category  
is about 35\%, and will not be counted in the muon-free category of 
multiplicity events (which corresponds to the main population of 
interest, i.e., high energy neutrons produced by muons in the rock).                                
The remaining 65\% of incoming neutrons will deposit less than 9\,MeV 
of prompt energy from the initial scatter followed by spallation of 
the Pb. When the prompt energy in these neutron-scatter-plus-multiplicity 
events is 3--8\,MeV, it will be indistinguishable from 
events without a neutron scatter but one unit higher multiplicity. 
For example, a multiplicity-three event with 7\,MeV of prompt energy 
will, to first approximation, appear the same as a multiplicity-four 
event with prompt energy below the 3-MeV threshold. Since both of 
these events are due to a high energy neutron, the inferred rate of 
high energy neutrons will not be biased.

Finally, muons that deposit less than 8--9\,MeV in the scintillator
(or none at all) but interact in the Pb and cause multiplicity events,
represent a potential background to the multiplicity events
due to high-energy neutrons. Of the estimated 350\,muons/day that 
will pass through the Pb, there could be a few per day that cause such
an event. However, these could be vetoed with a simple set of veto
counters placed below the lead, and used in anticoincidence. 

In summary, our design studies show that an apparatus consisting of a 
Pb target of 200\,cm by 200\,cm area by 60\,cm thickness covered by a 
60-cm-thick scintillation detector with Gd-capture detection 
efficiency of $\varepsilon_s(T)$, where $T$ is the low energy threshold 
for each distinct capture, and assuming an efficiency to detect 3\,MeV gamma-rays
in the Gd-scintillator tanks close to 100\%, 
will yield a rate for $M$-fold multiplicity-tagged events of 
$$R = N\,(1-0.35)\,(\varepsilon_s(T))^M\,\hbox{events/year},$$
where $N$ is the number of high-energy neutrons that induce an event
with $M$ or more detectable neutrons emerging from the Pb and entering
the scintillator, and the factor of $(1-0.35)$ is due to neutron
interactions in the scintillator that exceed the high energy
threshold. Our {\small FLUKA} and {\small MCNP}-PoliMi
simulations indicate that $M$=3 gives $N$=255 and
$T$=3\,MeV gives $\varepsilon_s(T)$=0.75, and therefore
$R=70\pm8$\,events/year. Depending on the actual gamma rate and
spectrum, some optimization is possible for increasing $R$ but protecting
against random multiplicity events, for example, by increasing the
multiplicity requirement and lowering the energy threshold. Generally
speaking, our method is capable of measuring the rate of high-energy
neutrons to about 12\% statistical error in the span of a year at a depth 
of 2000\,m.w.e. The expected number of background events, which is dominated 
by the rate of random gamma-induced coincidences,  is expected to 
be at most 10 events/year, and could be further suppressed by optimizing the 
multiplicity and energy thresholds. Note that  we are not trying to deconvolve the primary neutron energy
based on the detectable multiplicity shown in Figure~\ref{fig:anal_multiplicity_60cm}. 
Nevertheless, we can set a threshold on the primary neutron energy 
because unvetoed events with a multiplicity of 3 or more will be neutrons with a minimum 
energy of $\sim$60\,MeV given that lower energy ``primary" neutrons will scatter with protons 
in the scintillator and reach the Pb target with much lower energy.

\section{Discussion of other potential applications within shielding for Underground Dark Matter Experiments}
\label{statistical_and_veto}

In this section we present an application of a neutron multiplicity meter detector 
to a running dark matter experiment that serves as both an active shield 
and a monitor of the presence or rate of background events due to high 
energy neutrons. The idea exploits the same technique as a purpose-built 
instrument for background studies as described above. It's principal virtue in a  
\WIMP~search experiment is that it can closely monitor when a neutron  
background would appear in the data. A Gd-loaded liquid scintillator detector 
integrated into the shield would detect, using the multiplicity technique, the same  
population of events that cause a flux of low energy neutrons inside the  
shielded \WIMP~detector volume, namely, neutron multiplicity events  
produced in a Pb gamma shield by an otherwise undetected high energy  
neutron. Since the underlying processes are the same, Monte Carlo  
simulations would give a very reliable measure of the ratio of the  
rate of multiplicity events in the external detector to the rate of \WIMP-like 
events in the dark matter detectors due to the same neutron population.

Similar techniques to detect the presence of background sources have  
been successfully used, for example in the \CDMS-I~\cite{r21} and \CDMS-II~\cite{r119} 
experiments where multiple simultaneous nuclear-recoil events were used to determine 
the rate of single scatter nuclear recoils due to the same neutron background flux. 
The ratio of multiple nuclear recoil events to single nuclear recoils has the advantage 
of having a negligible source of systematic uncertainty since the neutron elastic cross 
sections on Ge and Si are very well known. Nevertheless, the rate of multiple nuclear 
recoil events is lower than the rate of single nuclear recoil events, and the uncertainty in 
the singles rate is dominated by the fluctuations of the multiples when only a small number 
has been observed.

In other words, tagged events that are correlated with the production of a 
single nuclear recoil due to a neutron can be used to statistically predict 
the absolute number of these nuclear recoils. If we call these tagged events 
``background predictors"  then the number of unvetoed singles can be estimated 
by determining the ratio of single nuclear recoils to the background 
predictor events with a Monte Carlo simulation, and then counting the 
number of background-predictor events in the experiment. Narrowing the 
statistical and systematic uncertainty of this ratio improves the ability to 
monitor and subtract the neutron background. Dark matter experiments that have a Pb 
layer or any high-$Z$ material as their gamma shield could use an external 
multiplicity meter to predict, in a statistical way,  the 
number of unvetoed nuclear recoils due to neutrons. The background 
predictor events with high multiplicity are detected in the multiplicity meter 
outside the high-$Z$ material. A virtue of this configuration is that the gamma 
background due to contaminants in the scintillator and Gd are shielded by 
the high-$Z$ material.

\begin{figure}[htbp]
\begin{center}
\includegraphics[width=4in]{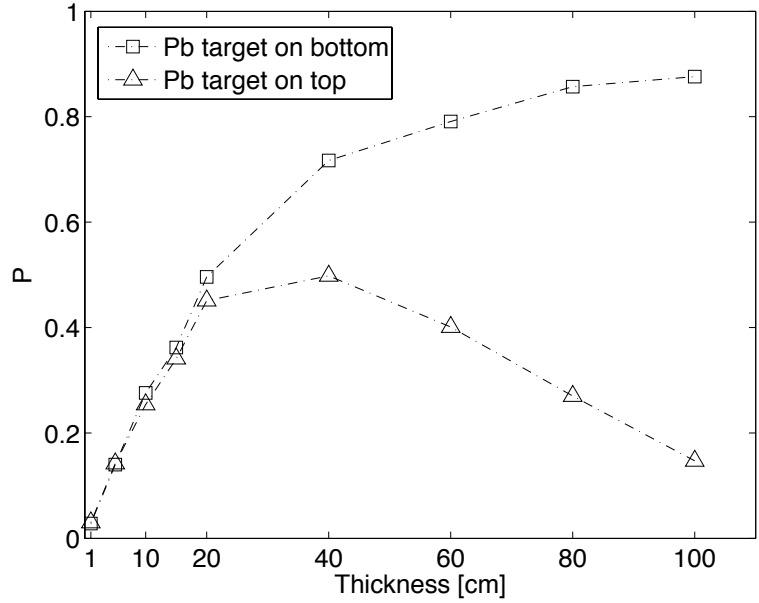}
\caption{\small Simulation with FLUKA to explore optimal target
thickness and position of the Gd-loaded scintillator tank with respect
to the Pb target. The parameter $P$ is defined as the fraction of
events (relative to the number of 100-MeV incident neutrons) that has 3
or more neutrons of 10\,MeV or less going towards the top or the
bottom of the Pb target. (See text for details.) Since a given event
may have 3 or more neutrons going to the top and 3 or more going to
the bottom, it is possible to have $P_{\hbox{TOP}}+P_{\hbox{BOTTOM}}>1$,
for example as observed for 40-, 60- and 80-cm thickness.}
\label{fig:flipped_or_NOT}
\end{center}
\end{figure}

\begin{figure}[htbp]
\begin{center}
\includegraphics[width=4in]{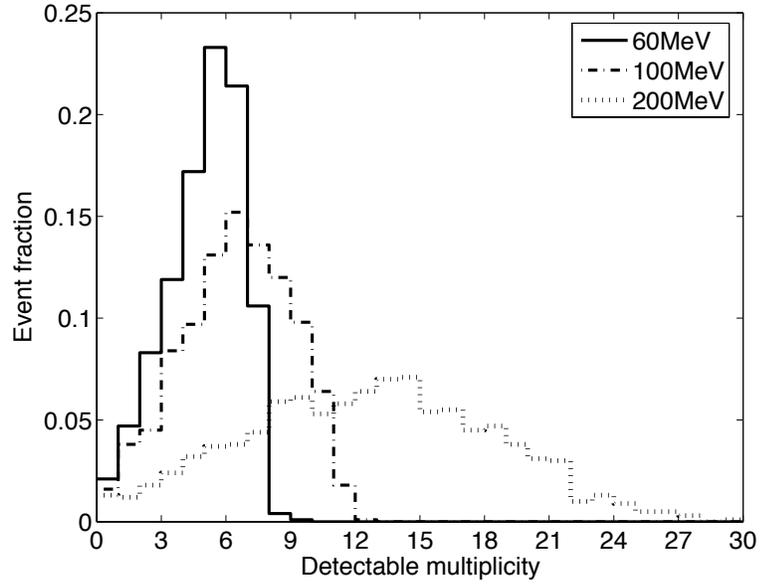}
\caption{\small A FLUKA simulation was done with a fixed target
thickness of 100-cm and varying the incident neutron beam energy in
order to explore the correlations between the energy of the incident
high-energy neutron on the target and the detectable multiplicity. If
we reference the 
beam direction a ``downward," the detectable multiplicity is
determined by counting the neutrons that reach a surface just above
the Pb target.}
\label{fig:mult_dist}
\end{center}
\end{figure}

\begin{figure}[http]
\begin{center}
\includegraphics[width=4in]{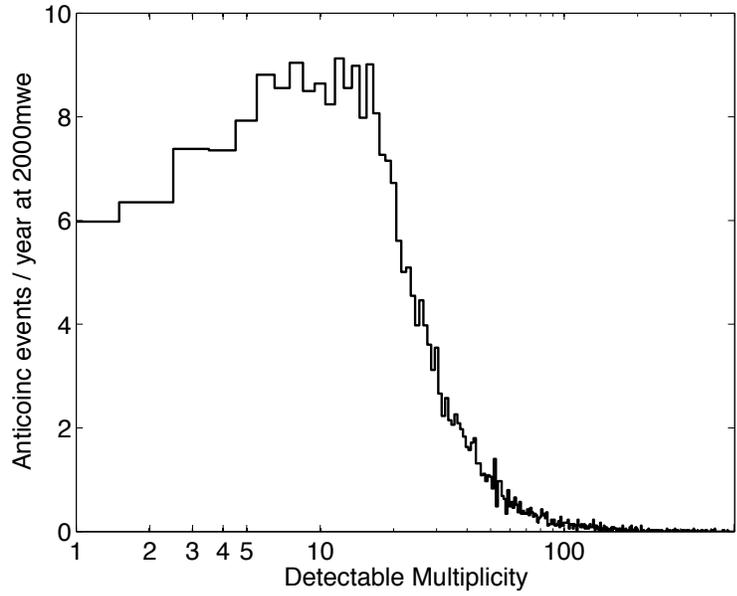}
\caption{\small The detectable multiplicity from a Pb slab of 200\,cm
by 200\,cm area and 60\,cm thickness for the events estimated to
be anticoincident with an energy deposition of 2\,MeV or more from
charged particles, including muons and hadrons. The detectable
multiplicity was counted only by looking at neutrons with energy less
than 10\,MeV going towards a surface on top of the Pb target.}
\label{fig:anal_multiplicity_60cm}
\end{center}
\end{figure}

\begin{figure}[htbp]
\begin{center}
\includegraphics[width=4in]{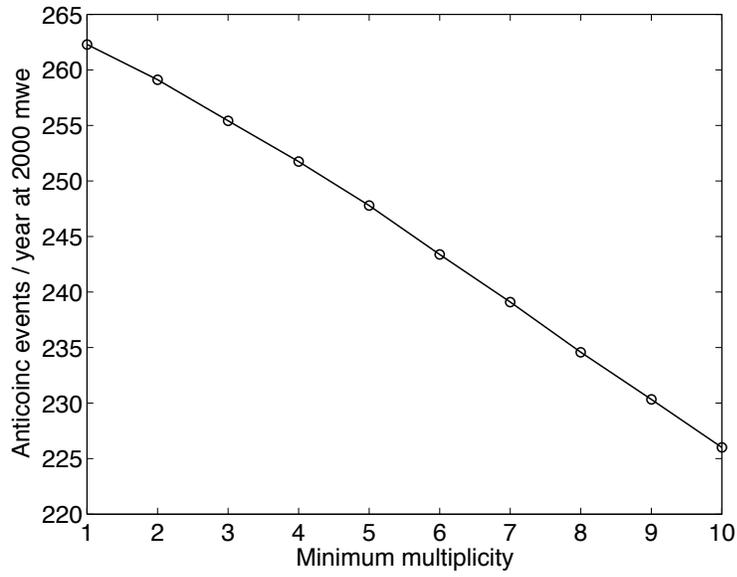}
\caption{\small The total number of events as a function of 
minimum multiplicity. The total number of events changes only by 
about 10\% between a minimum multiplicity of 3 and 10.
}
\label{fig:anal_effmult}
\end{center}
\end{figure}

\begin{figure}[htbp]
\begin{center}
\includegraphics[width=4in]{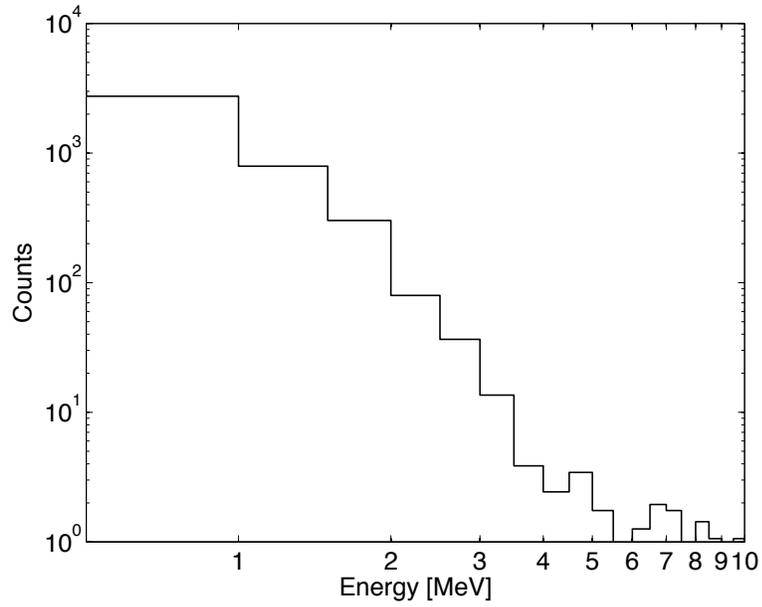}
\caption{\small 
Energy spectrum of secondary neutrons produced by high-energy neutrons
(flux shown in Fig.~\ref{fig:hencosz}) incident on the Pb
target. Neutrons mostly have energy below 5\,MeV energy, and indicates
that the thickness of the scintillator is not
driven by the moderation requirements. Rather, we find that the
thickness is driven by the need to efficiently contain the gammas
emitted by the Gd.}
\label{fig:anal_fluxspec}
\end{center}
\end{figure}

\begin{figure}[htbp]
\begin{center}
\includegraphics[width=4.5in]{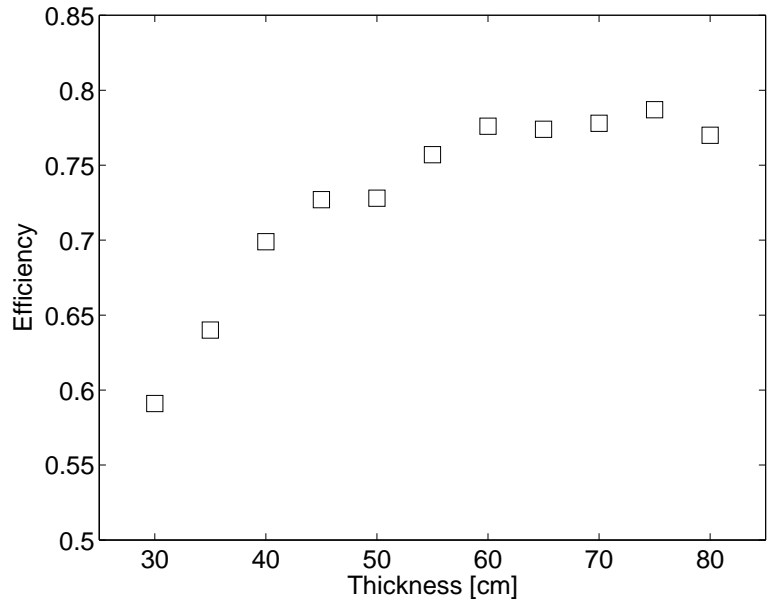}
\caption{\small 
Simulation with MCNP-PoliMi~\cite{mcnp-polimi} of the Pb target with Gd-loaded
scintillator contained in tanks placed on top of the target.  A beam
of neutrons with energy 0.5\,MeV was propagated from the Pb to the scintillator
tank. The thickness of the scintillator tank was varied. Efficiency corresponds 
to the fraction of incident neutrons for which the energy deposited in the  Gd-loaded 
scintillator by gamma rays is above 3\,MeV.}
\label{fig:scint_thickness}
\end{center}
\end{figure}

\begin{figure}[htbp]
\begin{center}
\includegraphics[width=4.5in]{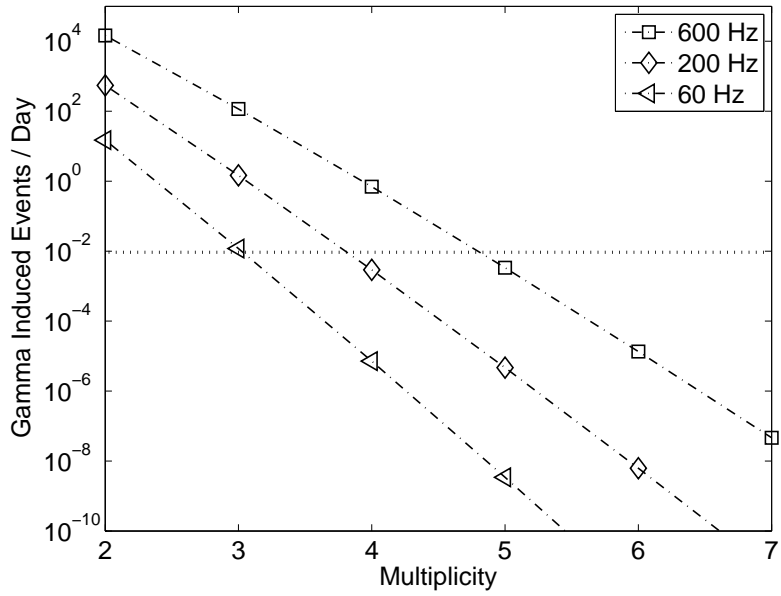}
\caption{\small Ambient gammas can mimic a high energy neutron event
due to accidental coincidences. The rate of gamma-induced background
events is plotted as a function of the multiplicity of the events for
a time window of 40 microseconds and three different gamma
rates~\cite{SSEN}.}
\label{fig:gamma_bkg}
\end{center}
\end{figure}

As a further illustration based on Fig.~\ref{fig:flipped_or_NOT} for 100\,MeV incident 
neutrons and a typical 15-20\,cm thick gamma shield made of Pb, the fraction of high 
energy neutron events that produce multiple low energy neutrons going inside 
a Pb box will be roughly the same as outside. For example, the use of 
60-80\,cm  of Gd-loaded liquid scintillator outside the gamma shield 
layer allows the moderation of low energy neutrons originating from the 
radioactivity in the rock and at the same time functions as a neutron 
multiplicity counter that would allow the prediction of the number of 
neutron-induced events in the signal region. Note that the multiplicity threshold 
in this case should be set high enough so that gamma induced multiplicity events are kept 
at a negligible level, since the Pb layer would be about 20\,cm thick.

The effect plotted in Fig.~\ref{fig:flipped_or_NOT} shows that the low energy neutrons 
produced from the neutron spallation reaction can be detected by clean low energy 
neutron detectors inside the gamma-ray shield (for example with plastic scintillator) but 
we have also found that outside of the gamma-ray shield, a 60-80\,cm of Gd-loaded 
liquid scintillator with a threshold of a few MeV, would work as an active veto 
complementing the veto inside the gamma shield (or as a standalone veto depending 
on rejection requirements) and as a monitor of the muon-induced neutron background. 
Note that the thickness of the scintillator outside the gamma-shield is driven by trying to 
contain the gamma rays produced in the capture of a neutron by the Gd in order to have 
a high threshold to defeat ambient gammas from radioactivity, and also to keep moderating 
with high efficiency the neutrons produced from the radioactivity in the rock.

\section{Conclusions}
We have designed an instrument capable of measuring 
the high energy ($>$60\,MeV) muon-induced neutron flux deep 
underground. The instrument is based on applying the Gd-loaded 
liquid-scintillator technique to measure the rate and multiplicity of low energy 
neutron events, and exploiting that the dominant source of events with 
multiplicity greater or equal to 3 are high energy neutron interactions. 
This unique signature allows both for efficient tagging of neutron muliplicity 
events as well as rejection of random gamma backgrounds so effectively that 
typical low-background techniques are not required. These design 
studies based on Monte Carlo simulations show that an apparatus 
consisting of a Pb target of 200\,cm by 200\,cm area by 60\,cm thickness covered by a 
60\,cm thick Gd-loaded liquid scintillator (0.5$\%$ Gd content) detector 
could measure, at a depth of 2000\,m.w.e., a rate of $70\pm8$ (stat)\,events/year. 
 The expected number of background events is expected to be below 10 events/year
  and could be further suppressed by increasing the multiplicity threshold.
 We have discussed  the relevance of this technique to monitor the muon-induced neutron 
background in searches for dark matter in the form of  Weakly Interacting Massive 
Particles and, based on these studies, we also discuss the benefits of using a neutron 
multiplicity meter as a  component of active neutron shielding.

\section{Acknowledgments}
The authors thank the National Science Foundation
for financial support under grant numbers PHY-0503729 and AST-9978911. We also 
thank H.F. Nelson, M.R. Dragowsky, R.W. Schnee, S. Yellin, and D.R. Grant for useful 
discussions and comments. We also thank J.F. Beacom and V.A. Kudryavtsev for critical 
reading of the paper.

\newpage

\newpage

\newpage

\newpage

\newpage

\newpage

\newpage

\newpage

\newpage

\newpage

\end{document}